# Slocalization: Sub-µW Ultra Wideband Backscatter Localization

Pat Pannuto
University of California, Berkeley

Benjamin Kempke
University of Michigan

Prabal Dutta
University of California, Berkeley

## ABSTRACT

Ultra wideband technology has shown great promise for providing high-quality location estimation, even in complex indoor multipath environments, but existing ultra wideband systems require tens to hundreds of milliwatts during operation. Backscatter communication has demonstrated the viability of astonishingly low-power tags, but has thus far been restricted to narrowband systems with low localization resolution. The challenge to combining these complimentary technologies is that they share a compounding limitation, constrained transmit power. Regulations limit ultra wideband transmissions to just -41.3 dBm/MHz, and a backscatter device can only reflect the power it receives. The solution is long-term integration of this limited power, lifting the initially imperceptible signal out of the noise. This integration only works while the target is stationary. However, stationary describes the vast majority of objects, especially lost ones. With this insight, we design Slocalization, a sub-microwatt, decimeter-accurate localization system that opens a new tradeoff space in localization systems and realizes an energy, size, and cost point that invites the localization of every thing. To evaluate this concept, we implement an energy-harvesting Slocalization tag and find that Slocalization can recover ultra wideband backscatter in under fifteen minutes across thirty meters of space and localize tags with a mean 3D Euclidean error of only 30 cm.

## 1  INTRODUCTION

Classically, high fidelity localization has been restricted to devices capable of actively beaconing their position, placing an energy demand on the device to be localized, requiring large energy stores, and resulting in limited lifetimes. Recently, a body of work emerged that demonstrates the ability to locate passive RFID tags [30, 46, 48] or sufficiently large (i.e. human torso sized) tagless objects [2]. While the energy-free operation is appealing, these systems track their targets by observing changes in the environment, requiring that either the targets or their trackers move to be localized.

However, most things do not move. Indeed, a vast array of things from the TV remote to warehouse assets to deployed sensors can be considered "nomadic," stationary but for occasional migration [36]. A key corollary to this observation is that the update rate for tracking a nomadic object can be very low. To that end, this paper introduces *Slocalization*, a new localization system that can localize static tags in both static and non-static environments with decimeter-level accuracy for less than one microwatt. At this power level, Slocalization is suitable for use with the burgeoning array of batteryless, energy harvesting systems [4, 22]. A standalone Slocalization tag will well outlast the self-discharge lifetime of a standard coin cell battery [11, 32]. Slocalization achieves this ultra-low power budget by reducing the location update rate from order hertz to millihertz, or several minutes per location fix.

Slocalization lies at the intersection of two recent research thrusts: backscatter communication and ultra wideband (UWB) localization. Slocalization leverages backscatter to generate the UWB signals needed for high fidelity localization with minimal energy burden and utilizes the superior ranging resolution afforded by UWB signals to recover decimeter-accurate estimates of tag position. In contrast to prior UWB systems, Slocalization tags do not actively emit RF energy, they only reflect it, requiring a new system architecture to capture, decode, and make use of these signals.

One of the key challenges in backscatter communication is that RF path loss is suffered twice, as the tag is simply a passive reflector, resulting in very weak signals. FCC regulations further limit UWB signals to significantly lower energy than narrowband, yet with Slocalization we are interested in covering whole rooms. To inform design decisions and establish the feasibility of recovering signals, we develop a model for the UWB backscatter channel. We use this model to explore what kind of signal energy can be recovered and how one might go about leveraging long integrations of the channel over time to extract a backscattered signal.

To move UWB backscatter from theory to practice, we develop a bandstitched, integrating UWB transceiver architecture. Today, the only commercial UWB transceiver chip is the DecaWave DW1000. Unfortunately, this chip is tailored to 802.15.4a communications, providing a relatively high-level interface, and does not expose information on the underlying UWB channel to application developers. As both Adib [2] and Kempke [21] observe, developing a direct UWB frontend is prohibitively costly, requiring expensive or niche hardware. We extend Kempke's bandstitching receiver design to include transmission of UWB signals, demonstrating the first end-to-end bandstitched GHz UWB transceiver architecture.

At this point, the weak tag signals are in the noise and cannot be seen. To recover tag transmissions, Slocalization anchors integrate samples of the channel over time. As environmental noise is generally white and Gaussian, its integration over time will remain generally flat. Integration of the periodic signal from the tag will cause it to rise above this noise, so long as the tag's signal remains remains sufficiently stable during the course of the integration, that is, the tag has a good frequency source and does not move.

With UWB backscatter in hand, we introduce the Slocalization architecture, an overview of which is shown in Figure 1. Fixed anchors with known positions in an environment emit pulses to sound the channel impulse response. Slocalization tags use a backscattering technique to perturb the channel impulse response with a periodic signal. Anchors integrate repeated measurements of the channel to lift the tag signal above the noise. After sufficient integration to identify the backscattered signal, anchors compute the time offset between the arrival of the backscattered path and the direct line-of-sight peak from the transmitting anchor. These time difference of arrival estimates yield ellipsoids of possible tag locations for each pair of anchors. The Slocalization system finds the best intersection of these ellipsoids to realize tag position.

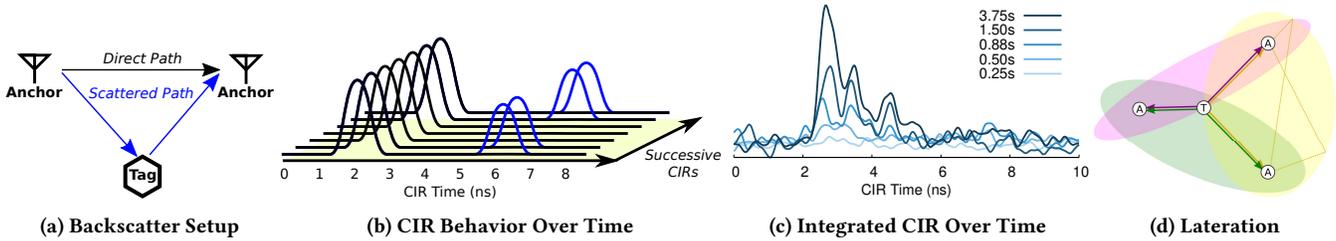

**Figure 1: Slocalization Concept of Operation.** (a) Anchors emit periodic pulses that sound the ultra wideband channel. A tag modulates its antenna to either reflect or absorb this signal. (b) perturbing the channel impulse response (CIR) over time. (c) Initially, the signal is too weak to detect. By integrating repeated estimates of the channel over time, the tag's arrival signal appears and its arrival time can be estimated. (d) Anchors use the time difference of arrival between the direct path between anchors and the backscatter path reflected from the tag to form ellipsoids of possible tag locations. The intersection of sufficient ellipsoids yields the absolute position of the tag.

To test whether the Slocalization system works in practice, we realize a prototype implementation. As we are motivated by the vision of a batteryless future, we design our Slocalization tag to be energy harvesting, including only a $5\,cm^2$ solar cell and a $47\,\mu F$ capacitor for transient energy storage to power the tag. With this tag and the Slocalization transceiver, we are able to demonstrate the recovery and localization of UWB backscattered signals.

Evaluating this prototype, we find that in a complex, indoor environment, Slocalization is able to localize the tag with only 30 cm average error. We evaluate the impact of varying the integration time on the quality of the Slocalization result, as well as the range of integration times required to localize a tag as distance increases. We then evaluate long-range performance, showing that across 30 m of space in both line-of-sight and non-line-of-sight conditions, Slocalization can estimate tag distance to within 0.1 m in under fifteen minutes. We show that Slocalization is robust to motion and other interference sources in the environment, and finish by establishing the viability of concurrently localizing multiple Slocalization tags.

In summary, the major contributions of this paper are the development of a decimeter-accurate, FCC-compliant localization system capable of localizing sub-microwatt, static tags in static or mobile environments; the introduction of the first ultra wideband backscatter platform; the presentation of a novel analysis of the ultra wideband backscatter channel; the development of a bandstitched ultra wideband transceiver architecture covering over one gigahertz of bandwidth; the introduction of integration to recover backscatter signals below the noise floor; and the demonstration of high-fidelity recovery of backscatter signals over thirty meters of free space in both line-of-sight and non-line-of-sight conditions.

## 2 BACKGROUND AND RELATED WORK

While the backscattering concept dates back decades [40, 45], there has been a recent resurgence in research around backscatter, extending the concept from beaconing simple identifiers to high bandwidth communication [44, 51], highly parallel communication [19], leveraging ambient environmental signals instead of active interrogators [25], or even motion capture [46, 48]. Localization is a similarly mature line of research, however, with the advent of new FCC regulations in 2002, the last decade has seen an explosion of interest in UWB for localization due to the greatly improved resolution it can provide indoors [8, 15, 27, 28].

Slocalization combines the best-in-class communication capabilities of backscatter with the best-in-class localization capabilities of UWB designs. We begin by reviewing these technologies and how recent progress in each subarea has informed and influenced the design of Slocalization.

### 2.1 Traditional Narrowband Backscatter

In traditional backscatter systems, an interrogator (e.g. an RFID reader) emits a powerful, well-known signal—often a pure sine tone. Tags in the environment modulate the impedance of their antenna by opening and closing a switch, changing their antenna from being highly reflective to highly absorptive. A receiver[1] captures these reflections and uses them to recover data from the tag. The key insight in backscatter is that it enables a vast energy asymmetry between the anchor (interrogator) and the tag, as the energy cost of actuating a switch to change impedance is very low.

### 2.2 Powering Backscatter Devices

Broadly, backscatter devices can be categorized as *passive* or *semi-passive*. A passive device ships with no local energy store, rather it opportunistically harvests energy from the RF signal of the interrogator. A typical energy budget for such harvesting is well below 1 mW, however projects such as the WISP [41] and the UMass Moo [50] have demonstrated that this is sufficient energy for an array of interesting computational applications. In contrast, semi-passive devices use an alternate power source, such as an on-board battery or indoor photovoltaics, for primary system power and leverage the RF channel solely for communication [5].

Under FCC regulations, narrowband readers can transmit up to 4 W EIRP (36 dBm), facilitating a 7-8 m operating range for classical RFID devices [5]. Unfortunately, the transmission power allotted for UWB devices is much lower, -41.3 dBm [14, 18]. Interestingly, recent work has demonstrated that it is possible to harvest as much as $16\,\mu W$ from a -18 dBm UHF signal, over 16× what is needed to power a Slocalization tag [38].[2] Our Slocalization prototype powers itself from a photovoltaic cell for simplicity, however any harvesting source (or energy store) capable of supplying $1\,\mu W$ can power Slocalization tags.

---

[1] In RFID, the interrogator (reader) is usually also the receiver, however Section 5.2 explores advantages and disadvantages of separating these roles.

[2] For a complete overview of modern RF harvesting, see Kim's summary [22].

## 2.3 Backscatter Channel Access

Mediating channel access is an interesting problem for the extremely limited budget afforded most backscatter devices. Ambient backscatter demonstrated that it is possible to develop a carrier sense mechanism that is tailored to the energy constraints of backscatter devices [25]. *Laissez-Faire* showed that for the transmission rates of backscatter, when communicating to a sufficiently capable receiver, one can simply ignore contention, transmit blindly, and let the receiver sort it out [19]. Directly adopting a laissez-faire approach would not work for Slocalization as our technique for recovering UWB signals would require unrealistically small jitter on the tags to preserve the subtle per-tag timing offsets used to distinguish tags. We do embrace tag simplicity, however. Slocalization requires no synchronization between tags and uses PN codes to distinguish transmissions from concurrently transmitting tags.

## 2.4 Localizing Passive Backscatter Devices

Classic RFID tracking does not precisely locate devices, rather it identifies which reader, if any, is nearest (via signal strength) [39, 43, 47]. Several research efforts have demonstrated true localization by examining the narrowband channel. RF-IDraw uses interferometry to trace trajectories, but can suffer from severe static offset of absolute position [46]. Others show that channel parameters can be used to recover more accurate positions, but these systems are limited to only a few meters range in practice [29, 48]. RFind sounds frequencies surrounding UHF RFID to further improve localization quality, but unfortunately is not FCC compliant[3] and still suffers the range limitations of other RFID systems [31]. RFly addresses the reader–tag range limitation using a drone as a powered (6 W) relay, but the drone must still travel to within a few meters of each tag [30]. In contrast, Slocalization achieves FCC-compliant, decimeter-accurate localization in whole rooms over 30 m in size.

## 2.5 Theoretical Systems

Some theoretical analyses explore the viability of UWB backscatter. As theoretical systems, these designs rely heavily on antenna and channel models to validate design choices. Unfortunately, the standard 802.15.4a channel model [34] is not well suited to modeling a "two-way" signal, i.e. a backscatter reflection, requiring simulations to mix in motion models or employ statistical tricks to attempt to model a complex, indoor UWB backscatter channel [17]. D'Errico et al. further explore how to design a hybrid system with a conventional RFID frontend for wakeup and energy harvesting [10]. The Slocalization design is independent of energy frontend and amenable to such a hybrid design.

## 2.6 Millihertz UWB Localization

The quintessential sensor networking technique to reduce energy consumption is to reduce duty cycle. If the argument is truly that devices rarely or never move, then perhaps running traditional localization systems at millihertz duty cycles is the right approach.

One immediate drawback for such a design is a poor peak to average power ratio, a prohibitive design point for battery-based systems. The capacitive storage banks of energy harvesting architectures, however, are well suited to intermittent high current operation. High peak power requirements do still require sufficient storage (in capacitor volume and board area) to support operations. To quantify these tradeoffs, we look at the state of the art in low power decimeter-accurate localization systems. For such a design, we only consider systems in which the underlying localization mechanism can achieve a stationary fix.

*2.6.1 Commercial Transceivers.* The lowest power decimeter-accurate single-fix localization with traditional radios is SurePoint, with 80 ms long ranging events at 280 mW, or 22.4 mJ per range [20]. SurePoint includes additional overhead to schedule and maintain time slots. However, for the sake of argument, let us assume that the very low duty cycle effectively eliminates interference and that there is zero static power draw between range events. To realize Slocalization's 1 μW, SurePoint can only range once every 6.2 hours.

For energy harvesting applications, SurePoint's 3.3 V operating level raises additional concerns. Using the harvesting and activation circuit from Monjolo [9], whose regulator is roughly 80% efficient across the 0.35-2 V input and 3.3 V/100-200 mA output range, requires 28 mJ in the storage capacitors, or roughly 14 cm$^2$ of board area for similar capacitors. The primary energy cost in SurePoint is the 145 mA DecaWave UWB transceiver. Even an order of magnitude improvement in transceiver energy would still realize only one transmission every 40 minutes at 1 μW.

*2.6.2 Impulse Frontends.* Prior implementations have also identified the transceiver as the most (energy) costly component and replaced it with a simpler and cheaper UWB pulse generator. The current lowest power decimeter-accurate, FCC compliant, single-fix localization system is Harmonium [21]. Capturing a location fix requires the tag to transmit for 53 ms at 75 mW, or 4 mJ per range. To realize a 1 μW average power budget, a Harmonium tag could transmit ranging pulses every 1.1 hours.

The Harmonium impulse generation circuit relies on exploiting the step recovery effect in RF BJTs. This requires the tag to have a relatively high operating voltage of 5 V. Again considering the Monjolo energy harvesting frontend, reaching 5 V adds an additional burden for energy harvesting designs. For a 5 V, 15 mA output, the regulator efficiency improves to 85% thus requiring 4.7 mJ in the storage capacitors, or 2.4 cm$^2$ of board area for energy storage.

A key aspect missing from the Harmonium system is differentiating multiple tags. The authors suggest having the tag modulate a PN code, where the code bit length is linearly proportional to the number of concurrent tags. However, this would result in a corresponding linear increase in the energy per range, resulting in a prohibitively energy-expensive transmission.

*2.6.3 Comparing Passive and Active Tags.* Ultimately, the energy required to open and close a switch (to reflect RF energy) is so much less than the energy required to radiate RF energy that even with a five order of magnitude increase in "transmission duration," backscatter consumes significantly less tag energy for a single location fix. These energy savings motivate exploring the viability of UWB backscatter-based localization.

---

[3]FCC 15.231(a) permits 12,500 μV/m only for control signals. The pure tones sent at each $f_s$ step do not qualify. Rather, RFind should be subject to the periodic limit 5,000 μV/m (or -21.2 dBm as opposed to -13.3 dBm). This reduces SNR to low single-digit values across the presented spectrum. However, RFind could leverage the integration technique presented in this work to recover sufficient signal—UHF Slocalization!

## 3 THE UWB BACKSCATTER CHANNEL

Backscattered signals are much weaker than those from an active transmitter as they must travel twice the distance. Recovering backscattered UWB signals is further confounded by limitations on UWB transmission power [14, 18]. The link budget for a Slocalization tag consists of three parts, also shown visually in Figure 2:

(1) Path loss from transmitter to tag
(2) Loss at the Slocalization tag
(3) Path loss from tag to receiver

The total combined path loss can be summarized through an adaptation of the Friis transmission equation:

$$P_r = P_t + G_t + G_{bt} + G_{br} + G_r +$$
$$20 \times \log_{10}\left(\frac{\lambda}{4\pi R_1}\right) + 20 \times \log_{10}\left(\frac{\lambda}{4\pi R_2}\right) - L_b \quad (1)$$

where $P_t$ and $P_r$ are the transmitted and received power, $G_t$ and $G_r$ are the anchor's transmit and receive gains, $G_{bt}$ and $G_{br}$ are the gains of the tag's antenna from the perspective of the transmit and receive antennas, $\lambda$ is the wavelength (in meters), $R_1$ and $R_2$ are the distances (in meters) between the tag and the receive and transmit anchors, respectively, and $L_b$ is the reflection loss ($2\times$ RF switch insertion loss). All gain and power figures are in decibels.

Using the example from Figure 2, with a (maximum permissible) transmitted signal power of -41.3 dBm/MHz and typical indoor settings of $G_t, G_{bt}, G_{br}, G_r$ = 0 dBi, $\lambda$ = 0.075 m, $L_b$ = 1 dB, and $R_1, R_2$ = 5 m, the power received from the backscatter tag is -159 dBm/MHz.

### 3.1 Integrating Signal from Noise

In a stationary environment with no other signal sources, the ambient noise is approximately white and Gaussian, that is its integral over a long period of time is roughly zero. This observation leads to the *slow* in Slocalization: namely if one integrates a sufficient number of samples over time, it is possible to extract the tag's signal from the channel impulse response. In Section 7.7 of our evaluation, we explore the impact of additional interference sources such as environmental motion or other ambient electronics, and show that these can be filtered out of the channel frequency response and do not significantly affect the performance of Slocalization.

Using the well-known interpretation of Johnson-Nyquist noise, we can express the noise as a function of integration time:

$$P_{dBm} = -174 + 10 \times \log_{10}\left(\frac{1}{t}\right) \quad (2)$$

where $P_{dBm}$ is the noise power and $t$ is the integration time in seconds. For intuition, integrating for 1 ms, 100 ms, 1 s, 1 min, or 1 h leads to noise of -144, -164, -174, -191, or -209 dBm respectively.

### 3.2 Integration Time vs Distance

Recall the goal is to measure the distance between the tag and an anchor by determining the time of arrival of the reflection from the tag. An SNR of approximately 26 dB in the channel impulse response is required for standard threshold-based leading edge detection techniques to accurately determine time of arrival [16]. From Equations (1) and (2), we should be able to derive a relation between anchor-tag-anchor distance and the required integration time.

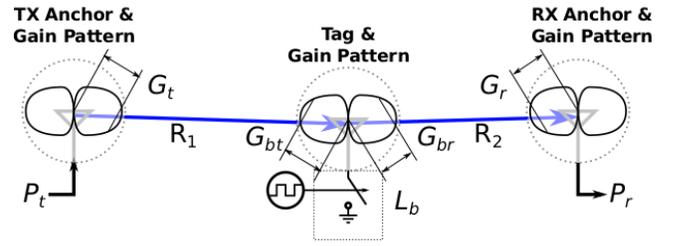

**Figure 2: Link Budget.** As the backscatter tag is not an active transmitter, its localization relies on the measurement of reflected signals from another active transmitting source. The recovered signal suffers path loss from the transmitter to the tag, losses internal to the tag, and path loss from the tag to the receiver. Slocalization requires long integration times to ameliorate these losses.

There are two small details we must address first. Equation (1) estimates the power at the receiver, however receive frontends also add noise, $\eta_r$, often around 10 dB in practice. Second, receivers directly measure the channel frequency response (CFR) to estimate the channel impulse response (CIR). As Section 4.2 explains, for a reasonable CFR resolution of 1,000 bins, coherent summation of integrated CFR samples will realize 30 dB of gain, $G_{CFR/CIR}$, in the CIR. Putting this together, we can express the required noise as:

$$\hat{P}_{dBm} = P_r - \eta_r + G_{CFR/CIR} - SNR \quad (3)$$

or $\hat{P}_{dBm}$ = -165 dBm for $R_1, R_2$ = 5 m and the typical values as used before. Substituting $\hat{P}_{dBm}$ for $P_{dBm}$ in Equation (2), it will require approximately 0.13 s of integration to recover the tag signal. More generally, using the estimates from this section, the minimum integration time required to recover the signal for a transmitter-tag distance $R_1$ and tag-receiver distance $R_2$ is:

$$t = 10^{-3.67} \times (R_1 R_2)^2 \quad (4)$$

A variety of factors including obstructing materials and nulls in the tag's antenna pattern can have a great effect on the parameters described in the backscatter path loss. Therefore, a significant margin of error must be applied in integration time to achieve high likelihood of tag detection in realistic indoor environments.

## 4 TRANSCEIVER DESIGN

The previous section described the UWB channel in theory. In this section, we explore the generation, manipulation, and recovery of backscattered UWB signals in practice.

### 4.1 UWB Bandstitching

To address the limited availability of UWB hardware, we previously presented the design of a bandstitched UWB receiver [21]. The idea of bandstitching is that a more traditional and accessible narrowband receiver can capture a UWB sample by taking a series of narrowband samples at successive frequencies (3.33-3.36 GHz, 3.36-3.39 GHz...), add these samples together in the frequency domain, and then use this "stitched"-together sample to recover a high-fidelity UWB channel impulse response in the time domain.

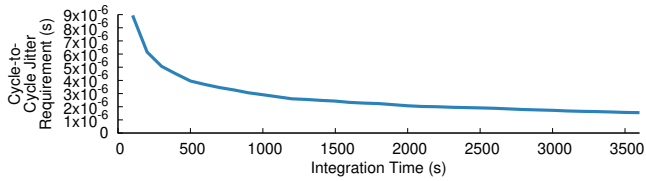

**Figure 3: Long Integrations Require Stable Crystals.** To recover tag signals, the receiver must be able to correlate the tag pulse train. This requires pulse generation to remain stable during the receiver's integration window. This curve (simulated for a 256 Hz tag frequency) shows how permissible tag jitter (phase noise) falls as the integration time increases.

We extend the principle to UWB transmissions, creating a bandstitched UWB transceiver. While this modification is fairly straightforward, bandstitching both the transmitter and receiver introduces an additional system-level constraint that frequency hopping between the transmitter and receiver must be synchronized. This is trivial for the monostatic case, where the transmitter and receiver are the same, but requires external synchronization for bistatic configurations (where transmitters and receivers are separated).

### 4.2 Backscatter Signal Recovery

Bandstitching captures the channel frequency response (CFR), but we are ultimately interested in using its dual, the channel impulse response (CIR), to estimate the arrival of the tag's signal. Recovery first requires searching for the precise tag frequency and phase offset, then integrating samples over time to enhance SNR, and finally estimating the arrival time of the tag signal.

**Signal Requirements.** To be able to extract the tag's signal, the tag's transmit sequence must have a zero mean, ensuring that no portion of the direct CIR is present after correlation. Additionally, the sequence must employ a modulation rate higher than that of other dynamic sources within the environment. Slocalization mixes the transmit sequence with a pattern of the form $\text{sgn}\left(\sin\left(2\pi f \times t\right)\right)$ to meet these requirements.

**Signal Stability.** Timing jitter in the tag's modulation sequence will cause the transmitted signal to shift slightly over time. To successfully recover the signal, over the course of the anchor's integration period, the modulation sequence must not deviate by more than 1/4 bit period from the average rate. Figure 3 shows the allowable signal jitter vs. integration time for the 256 Hz tag modulation rate used in this paper, derived through Monte Carlo simulation. One of the better available frequency sources, the AM0805, has an RC jitter of 500 ppm. While some research RC oscillators show promise towards tens of ppm [7], realizing the necessary stability with commercially available components requires the use of the higher-power crystal mode to maintain code coherence.

**Signal Discovery.** While the nominal frequency, $f = 256$ Hz, is known, in practice the frequency modulated by the tag may drift slightly, meaning the actual frequency transmitted will be some modest $\epsilon$ off the target. Furthermore, there will be a phase offset based on when the anchors begin sampling the CFR. This means that signal recovery must search the space $\sin\left(\left(2\pi f \times \epsilon_0\right) \times t + \phi_0\right)$ for the $\epsilon_0$ and $\phi_0$ that most strongly correlate, where $\epsilon_0$ is limited by the stability of the tag frequency source and $\phi_0 \in [0, \pi)$. This search introduces a system tradeoff explored in Figure 4. If the tag drifts more than half a cycle over an integration period, additional integration will begin destructively combining. Longer integration times require more precisely identifying the tag frequency, which increases the number of $f_{\text{cand}}$ that must be considered.

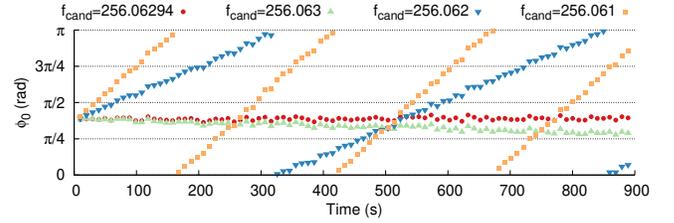

**Figure 4: Frequency Precision and Accuracy.** We record for 900 s with the tag near the anchor (so it can be found with short integration time). We break the recording into 10 s increments and search for the phase offset, $\phi_0$, for four fixed candidate frequency values, $f_{\text{cand}}$. Finding the precise frequency, $f_{\text{cand}} = 256.06294$, is computationally expensive. A coarser 0.01 Hz step exhibits low offset for $f_{\text{cand}} = 256.063$ over this sample. However, if we process this whole recording as one long integration, at about 500 s for $f_{\text{cand}} = 256.0622$, continuing to integrate would begin to reduce the recovered signal. With continuous integration, $f_{\text{cand}} = 256.0621$ would alternate between best possible and no signal roughly every 250 s when the tag is transmitting a simple square wave. Because of this, for signals that require long integration times to detect, if $f_{\text{cand}}$ is too far off, the tag will never be found.

**Signal Integration.** Integrating multiple samples over time is the key to pulling the tag signal above the noise floor. The actual integration is simple, just sum together all the correlated CFR estimates. Figure 5 shows the tradeoff between the number of CFR bins and the CIR variance. Due to the coherent summation of CFR bins, the required SNR for each CFR bin to realize a target CIR SNR decreases with an increasing number of bins. The coherent summation of $N$ bins yields a $10 \times \log_{10}(N)$ increase in CIR SNR. To achieve an approximate 26 dB CIR SNR[4] requires a CFR bin SNR of $26 - 10 \times \log_{10}(N)$, informing dwell time at each band.

**TDoA Estimation.** Once integrated, the individual bands can be stitched together in the frequency domain, and the inverse FFT yields the CIR. To find the TDoA, the arrival time of the direct CIR is subtracted from the arrival time of the tag's signal. Precisely estimating arrival time, particularly for lower SNR cases, is an active area of research [16, 52]. Our current implementation uses a simple thresholding approach. Section 8.4 explores how more advanced techniques could further improve Slocalization accuracy.

**Additional Tradeoffs.** The number of bandstitching steps along with the dwell time at each step defines the time to complete a full UWB sweep. Various methods can be employed to increase the UWB sweep rate. The instantaneous bandwidth can be increased through the use of higher sampling rate ADCs. Multiple bands can be observed simultaneously through observation across multiple center frequencies. Our prototype implementation employs 25 MHz of instantaneous bandwidth utilizing one RF receive frontend, yielding 49 steps to generate 1.225 GHz of UWB sweep bandwidth.

---

[4]26 dB of CIR SNR yields a negligible false positive rate in CIR ToA detection.

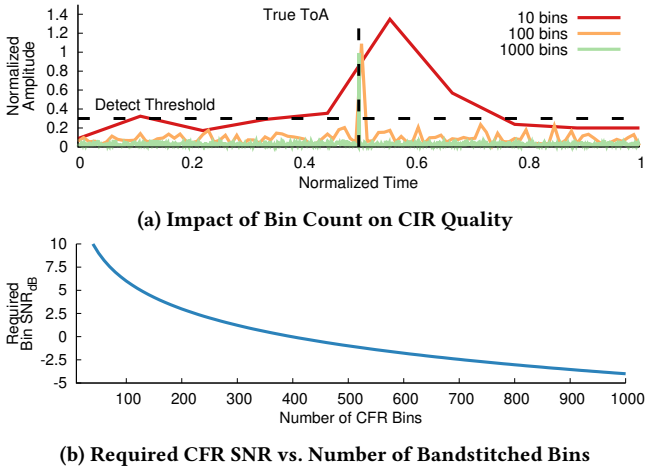

**(a) Impact of Bin Count on CIR Quality**

**(b) Required CFR SNR vs. Number of Bandstitched Bins**

**Figure 5: Processing Impacts Precision.** Introducing more band-stitching bins not only contributes to better CIR resolution from greater utilized bandwidth, but also improves the CIR SNR, given the same integration time for each CFR bin. The increase in SNR is due to the coherent contribution of many, noisy CFR bins. For the single-path case, the CIR SNR increases by $10 \times \log_{10}(N_{bins})$.

## 5 SLOCALIZATION DESIGN

In the Slocalization architecture, a localization event begins with a network of infrastructure nodes sounding the UWB channel. UWB reflectors in the space appear as perturbations in the channel impulse response (CIR) recovered by the infrastructure nodes. A tag in the environment opens and shorts its antenna such that one such reflection appears and disappears reliably over time. By comparing the difference between the direct, line-of-sight (LoS) path and the tag's backscattered path, a pair of infrastructure nodes can determine an ellipsoid of possible tag locations. With sufficient infrastructure nodes, the intersection of ellipsoids reveals the tag's final location.

### 5.1 CIR Perturbation (Tag Design)

Conceptually, a Slocalization tag is very simple. Figure 6 shows the complete architecture. The energy source could be an energy harvesting frontend or simply a battery. As discussed in Section 4.2, all a tag needs to do is toggle an RF switch at a stable frequency. To distinguish multiple tags, Slocalization inserts a cyclic shift register holding a PN code between the oscillator and the RF frontend.

### 5.2 CIR Coverage (Anchor Placement)

To localize tags, Slocalization anchors must capture estimates of the time of flight from an anchor, to a tag, to an anchor. One key question is whether the transmitting and receiving anchors should be the same—a *monostatic* configuration—or separated in space—a *bistatic* configuration. Recall that the distance from the anchor to tag to anchor traces out an ellipsoid of possible tag locations, with the anchors as the foci. In a monostatic configuration, the foci are overlapped, creating a sphere of possible tag locations.

In practice, these different shapes will change the best, average, and worst case integration time across space in an environment.

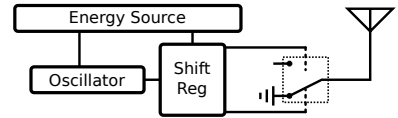

**Figure 6: UWB Backscatter Tag Design.** A UWB antenna and RF switch are used in conjunction to modulate the reflective characteristics of the RF channel. A shift register stores a PN code for the tag to emit. A high-stability oscillator clocks the shift register to drive backscattered communication.

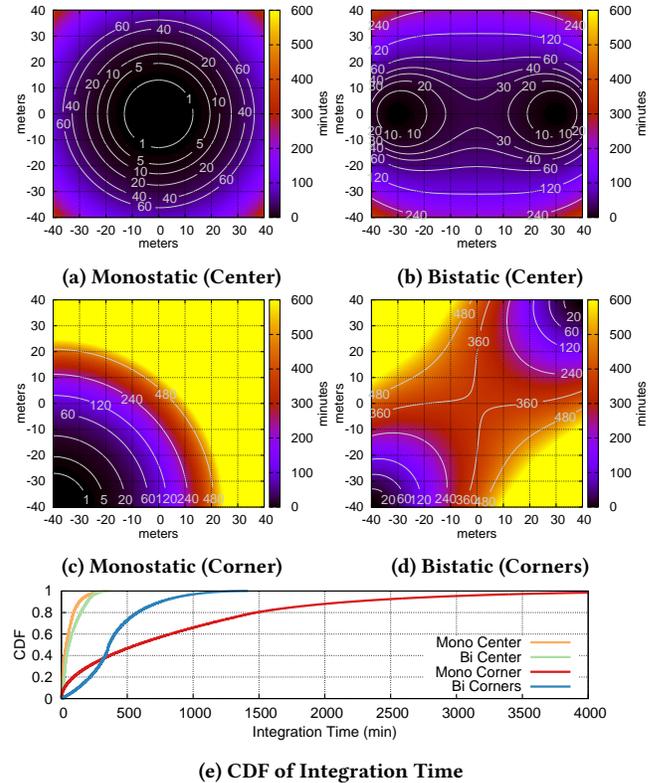

**(a) Monostatic (Center)**  **(b) Bistatic (Center)**

**(c) Monostatic (Corner)**  **(d) Bistatic (Corners)**

**(e) CDF of Integration Time**

**Figure 7: Anchor Arrangement Affects Integration Time.** The transmitting anchor can either be co-located (monostatic) or separated from the receiving anchor (bistatic). Monostatic arrangements suffer from high flash amplitude (the limited dynamic range of the RF frontend is overwhelmed by nearby high energy reflections) and inadequate spatial coverage in large areas. Bistatic results in a better coverage but requires time synchronization between the transmitting and receiving anchors, now physically separate.

Figure 7 considers four possible two-anchor placements for an $80 \times 80$ m room: first placing anchors for the best case monostatic and bistatic coverage and then a more realistic scenario with anchors mounted in corners of the room. While the ideally placed monostatic setup achieves the best coverage, it is unreasonable to expect an anchor to be placed in the center of every room. For the more realistic corner-based deployment, the bistatic configuration performs much better in the medium and long tail. For this reason, we use a bistatic anchor configuration in our implementation.

### 5.3 CIR Measurement (Anchor Coordination)

While Section 4.2 covers the signal processing to recover a distance estimate, Slocalization also requires that anchors coordinate so as not to trample each others' channel estimates. Furthermore, in a bistatic configuration, Slocalization anchors must also synchronize the bandstitching steps between transmitter and receiver.

To reduce implementation complexity, Slocalization follows in the footsteps of WiTrack and Harmonium and simply runs a wired sync pulse to all of the anchors. We note that several potential methods for accurate decentralized time synchronization have been explored in previous work using both wireless [12, 33] and wired techniques [13, 26], and leave their integration for future work.

## 6 IMPLEMENTATION

All software and hardware designs are open source and made available to the research community at github.com/lab11/slocalization.

### 6.1 Hardware

Implementing Slocalization does not require many components. However, due the sensitivity of the backscatter channel and a focus on minimal power draw, careful selection of components is required to maximize the potential of Slocalization.

**The tag**, shown in Figure 8, uses the UPG2422TK RF switch due to its minimal insertion loss, low power operation, and low switching voltage. An MCU emulates the functionality of a shift register and is used to facilitate greater experimental flexibility. To allow deepest sleep, the RF switch control lines are held by flip flops and the frequency reference provided by a 50 nA RTC. The energy frontend consists of an indoor photovoltaic cell and a low-leakage capacitor.

**Anchors** are USRP N210s synchronized with a shared clock and connected via gigabit Ethernet to a host computer that coordinates bandstitching. Transmit data are fed to the designated TX anchor as a repeating sequence of twenty IQ samples, chosen as a sequence that minimizes dynamic range and maintains equal amplitude across the 25 MHz of bandwidth occupied at each step. Due to the repetitive nature of the signal, this sequence is designed to generate twenty CFR peaks across 25 MHz, calibrated to a transmit amplitude abiding by the FCC requirement of -41.3 dBm/MHz.

Receivers feed IQ samples back to the host PC for post-processing. An initial real-time integration step averages out high frequency effects.[5] The 20-sample sequence is integrated one thousand times before offloading the averaged IQ data. This 1000× decimation yields a CFR update rate of 1.25 kHz, enough to cover the Slocalization modulation rates while minimizing signal processing complexity.

### 6.2 Processing

All processing is performed in MATLAB on raw USRP data.

**Data Parsing and Trimming.** Averaged IQ data includes tagged metadata identifying the precise time and target of retune events, which are used to segment the IQ data into separate bandstitching snapshots. After IQ data segmentation, the first 80 ms of each step are trimmed to allow the receiver's RF PLL to settle to the newly-tuned frequency.

---

[5]At 20 samples/repetition and 25 Ms/s, a CFR update rate of 1.25 MHz is achievable but not useful for Slocalization's low tag modulation rates.

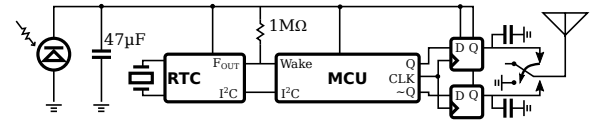

**(a) Tag Schematic**

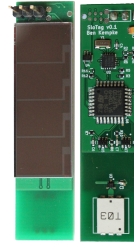

| Part | MPN | Quantity | Cost (USD@1k) |
|------|-----|----------|---------------|
| MCU | STM32L051K8T6 | 1 | $1.80 |
| Antenna | AH086M555003-T | 1 | $1.57 |
| Solar Cell | AM-1417 | 1 | $1.44 |
| RF Switch | UPG2422TK | 1 | $0.71 |
| RTC | AM0805AQ | 1 | $0.55 |
| Crystal | AB5OT-32.768KHZ-7-T | 1 | $0.38 |
| Flip Flop | 74LVC1617S | 2 | $0.09 |
| Passives | — | — | $0.16 |
| PCB | — | 1 | $1.00 |
| Total | | | $7.70 |

**(b) Tag**        **(c) Bill of Materials**

**Figure 8: Realized Tag.** We insert a low-power MCU in place of a shift register for flexibility. We use an ultra low power real time clock from Ambiq to achieve the requisite oscillator stability for minimal power. To minimize active power, we sleep the MCU between (potential) bit flips, requiring a pair of flip flops to drive the RF switch. The tag is powered with a small (3.5 cm × 1.4 cm) solar cell and limited energy storage (47 µF) to demonstrate its applicability to demanding energy harvesting applications.

**Clock Ambiguity Resolution.** Time is distributed as a 10 MHz signal to each anchor, which multiplies it 10× to provide clocking internal to the USRP. This reference is then divided by 4× to provide the reference for the transmit/receive RF PLL. Depending on the random timing introduced through the power-on sequencing internal to each radio, the phase of the final 25 MHz signal can be offset in time between anchors. A signal processing step in software measures the phase difference incurred between received bands and corrects for any phase offset incurred.

**Tag Frequency and Phase Search.** Our implementation searches for a nominal frequency of 256 Hz ±500 ppm in 5 ppm steps and eight possible phase offsets at each step. Each candidate is fed through a Blackman window and the {frequency, offset} pair with the strongest correlation is selected.

**Integration and Calibration.** Next, correlated CFR samples are integrated (summed in time). A one-time calibration performed in advance captures pairwise recordings of direct connections between each pair of anchors. To compensate for any phase offset incurred during RF signal generation and reconstruction, the integrated CFR is deconvolved with the calibration data to yield the actual CFR.

**TDoA Estimation.** The direct CFR is recovered by stitching the captured CFRs with no correlation step and then deconvolving with the calibration data. To improve the resolution of the CIR, the CFR is zero-padded to be 10× longer before applying the Inverse Fourier Transform. To estimate signal arrival time, we use the 30% height of the tallest peak in the CIR. The TDoA estimate is the difference in ToA between the direct and backscatter CIRs.

**Localization.** TDoAs between a tag and participating anchors define ellipsoids of possible locations. A minimum mean squared error solver uses gradient descent to find a best-fit position estimate.

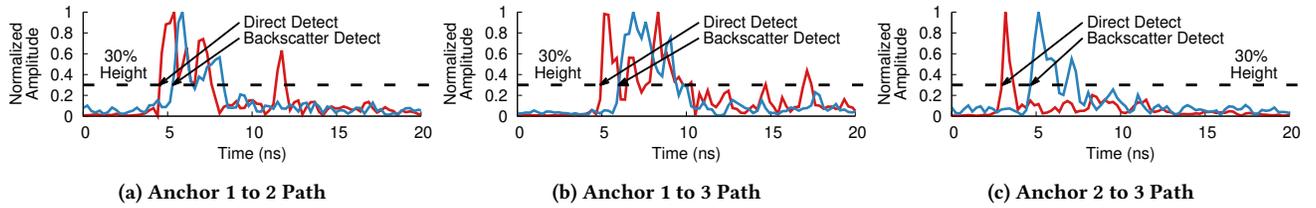

**Figure 9: TDoA in the Channel Impulse Response.** CIRs estimated from 1.225 GHz of bandstitched narrowband measurements for three anchor pairs. The difference in time between the direct line-of-sight measurement and the backscattered signal yields the distance between the tag and anchors. Multiple anchors with a TDoA measurement from each are necessary to determine a tag's 3D location accurately.

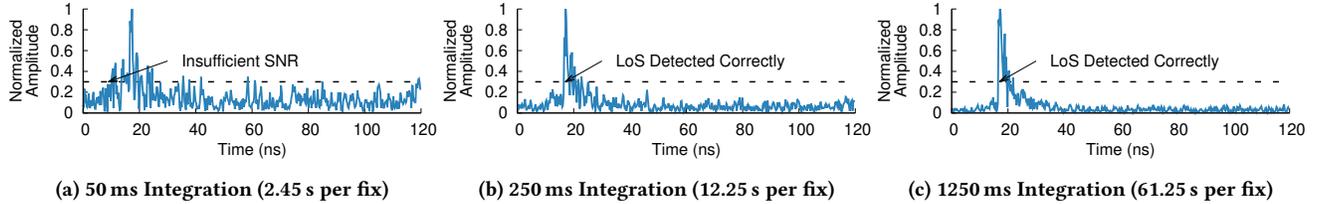

**Figure 10: Effect of Integration Time on Channel Impulse Response and Arrival Time Estimation.** The 30% height of the CIR's leading edge is used to estimate the arrival time of the line-of-sight path, necessitating sufficient SNR to resolve the leading edge. A number of integration lengths are shown for an example backscatter CIR. While 50 ms of integration time exhibits insufficient SNR to resolve the line-of-sight path, anything more than 250 ms shows sufficient SNR to resolve the backscatter CIR in this link scenario.

## 7 EVALUATION

We aim to establish the viability of Slocalization and explore its potential. We demonstrate recovery of TDoA estimates from a backscatter signal, explore the impact of varying integration, and evaluate end-to-end localization performance, finding Slocalization achieves 30 cm average error across an array of points. Then, we evaluate the long range—and long integration—performance by localizing a tag between anchors that are 30 m apart, first under direct line-of-sight and then non-line-of-sight conditions. We next evaluate some of the underlying Slocalization components and investigate how Slocalization can handle and reject environmental interference. Finally, we show that we can distinguish and recover ranging information from multiple Slocalization tags transmitting in parallel in the same environment.

### 7.1 Can Slocalization Measure TDoA?

We set up three anchors configured for bistatic ranging and a single tag. Figure 9 shows the recovered CIR for the Anchor 1 → 2, 1 → 3, and 2 → 3 paths. The Slocalization system can clearly identify peaks for both the direct and backscattered path for all anchor pairings. This time difference of arrival (TDoA) coupled with known 3D positions of anchors can be used to localize the tag.

### 7.2 Integration Time

Integration time is the key factor that determines how fast Slocalization runs. Because the signal received from the tag is well below the noise floor, the Slocalization system needs to integrate numerous samples of the environment over time to extract the tag's signal. Recall, the goal is to be able to accurately detect the leading edge of the pulse reflected by the tag, as the time offset of this edge yields the distance between the tag and anchors. Figure 10 looks at the effect of varying this integration time for a sample link.

For this experiment, the anchor-tag-anchor distance is just shy of 5 m, which allows us to push integration time down to 250 ms and still successfully recover the line-of-sight path. Note that 250 ms is only the integration time for one slice of the UWB spectrum. Bandstitching requires 250 ms of dwell time at each of the 49 frequency slices, thus requiring 12.25 s to fully resolve position.

### 7.3 3D Location Estimation

We next investigate the quality of the location estimates provided by Slocalization. We set up Slocalization in a 4.5 m × 3 m × 2.3 m indoor space—the room is typically furnished with tables, chairs, cabinets, etc., but with line-of-sight paths available between the tag and each anchor—and place the tag in 10 different locations on a table in the room. We configure the bandstitching sweep to dwell for 2 s at each of the 49 measured bands, requiring 98 s total for each location fix, an update rate of approximately 10 mHz. Figure 11 shows the estimate and ground truth of a single location fix at 10 points in space and finds that the Slocalization system is able to achieve an average error of only 30 cm across all 10 locations.

### 7.4 Long-Range Performance

A key differentiator of Slocalization from prior RFID-based localization systems is the ability to cover large areas. To evaluate this, we place two anchors 30 m apart in a long hallway. We set the tag 1 m away from anchor A (29 m from anchor B) and move it at 1 m increments to the center point (15 m from each anchor), as shown in Figure 12a. We configure Slocalization to dwell for 20 s at each band, recording 16.3 min of data at each location. Each point captures two measurements, swapping the transmitter and receiver role among the anchors. This experiment runs for over eight hours, during which people move through the evaluation space (a hallway connecting occupied offices) normally.

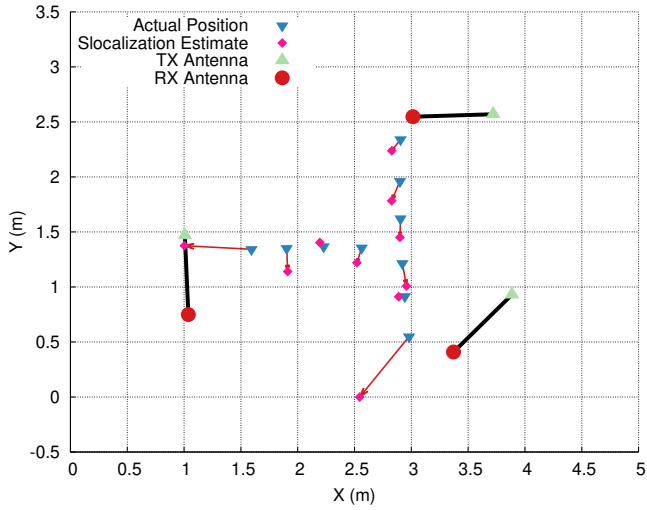

| | Min | Max | Mean | Median |
|---|---|---|---|---|
| 2D XY Euclidean Error (m) | 0.05 | 0.70 | 0.25 | 0.18 |
| 3D Euclidean Error (m) | 0.08 | 0.70 | 0.30 | 0.26 |

**Figure 11: Slocalization Performance Evaluation.** Ground truth vs. estimated tag position in a 4.5 m × 3.0 m × 2.3 m interior room. A number of fixed locations are chosen for the Slocalization tag, and the difference between the calculated position and the true position are shown. Slocalization is able to achieve 30 cm of average 3D error using sub-microwatt tags across the entire evaluation space using only 98 seconds of integration time at each location.

We iteratively feed progressively longer samples of the data into the Slocalization TDoA estimator, checking the result against the expected TDoA and reporting when the estimate reaches accuracy targets from 0.1 m to 5 m. Full results are shown in Figure 12b. At the center point, furthest from each anchor and thus requiring the most time, Slocalization requires 18 s of integration per band, or 14.7 min total, to localize the tag to 0.09 m error. Manual examination of the data around the 12 m data point reveals that the tag's signal was eventually recovered, but both the backscatter and the direct CIRs are ambiguous. Around this time, a small crowd of people carried a conversation directly in front of anchor B. While there is some reliability to non-line-of-sight conditions, UWB signals cannot reliably penetrate multiple bodies and travel 30 m.

## 7.5 Nulls and Reliability

Our prior work in UWB localization has shown that UWB channel robustness is greatly enhanced by incorporating multiple antennas at each anchor, ideally three at 120° offsets [20]. Our Slocalization prototype does not realize full antenna diversity. Rather each anchor simply has one dedicated transmit antenna and one receive antenna, separated by 72 cm. Figures 12c and 12d break apart the previous experiment, showing the performance of each path. While the exact cause of failures, such as the 9 m point in either direction or the longer ranges for A→B, can be hard to ascertain, greater path diversity, such as recording on both antennas while acting as the receiving anchor, would improve Slocalization robustness.

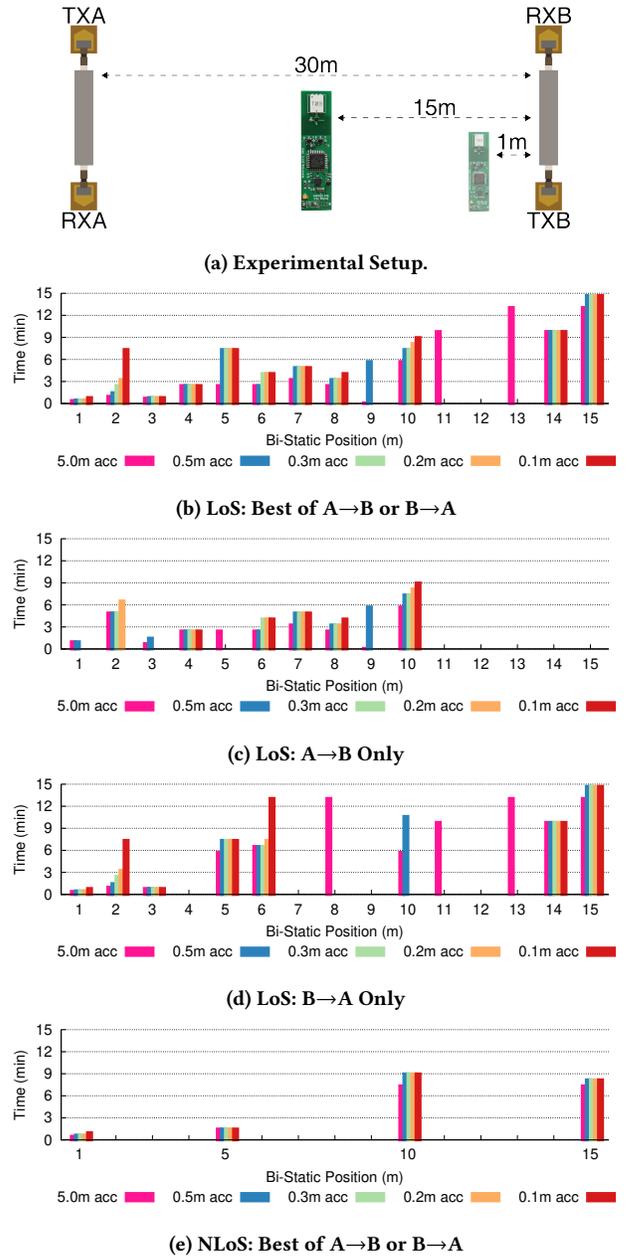

**Figure 12: Long Range and NLoS Performance.** We set up two anchors 30 m apart in a long hallway. We place the tag at 1 m increments, moving from anchor A towards the center of the hallway. For each location, we configure each anchor to both transmit and receive, collecting 20 s of integration per band, or 33 min per location. We iteratively process each sample to find the minimum integration necessary to reach varying accuracy targets, finding Slocalization requires only 14.7 min for the worst-case 15 m position. We then simulate an "in-walls" deployment by occluding both anchors with large tiles and measuring the NLoS performance at 5 m steps, finding that Slocalization performs better in this case. With anchors in the corners, Slocalization could localize an entire 15 m × 15 m room to decimeter accuracy in under fifteen minutes.

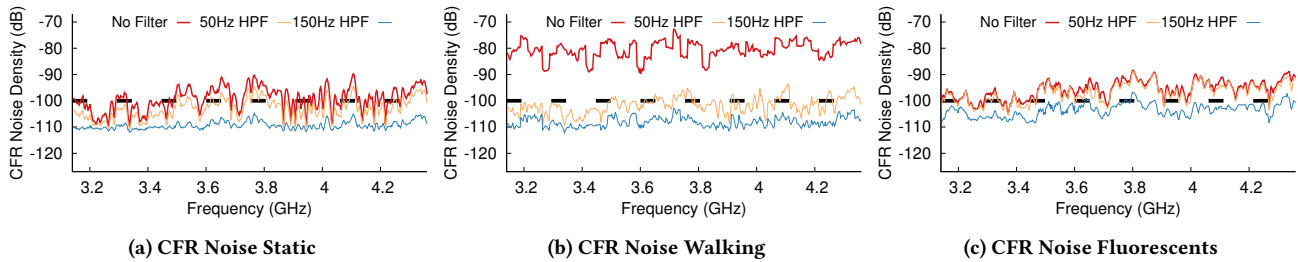

(a) CFR Noise Static

(b) CFR Noise Walking

(c) CFR Noise Fluorescents

**Figure 13: Effects of Dynamic Environmental Processes on CFR.** Slocalization must compensate for dynamic changes in the environment to be able to detect backscattered signals. Here we see the effects of different dynamic channel conditions on the CFR, the noise it imparts, and the effect of various filtering strategies. The dashed line is the required noise density requirement of a typical backscatter link with 100 dB of path loss. Walking around the environment imparts low-frequency noise which can be easily compensated through the use of a 50 Hz high-pass filter on CFR observations. Dynamic changes due to fluorescent lighting imparts higher frequency noise, requiring the use of a higher frequency high-pass filter to cancel. A control run shown in (a) shows that even seemingly stationary environments observe CFR noise, likely due to noise internal to the software-defined radio. To minimize active power, the tag should set its modulation rate as low as possible, however these effects require setting the modulation high enough to not be drowned out by these common sources of noise. The chosen 256 Hz modulation rate balances these tensions.

## 7.6 Non-Line-of-Sight

Real-world deployments may wish to hide infrastructure nodes. To simulate "in-wall" anchors, we place a $0.6 \times 1.2\,\text{m}$ tile in front of each anchor and re-run the experiment from Figure 12a placing the tag at the 1 m, 5 m, 10 m and 15 m positions, with results in Figure 12e. Somewhat surprisingly, the NLoS performs better, needing only 8.2 min to localize the tag to 0.1 m accuracy at the 15 m point. Qualitatively, the recovered backscatter CIRs look smoother and less noisy from the NLoS experiments, suggesting that the obstruction perhaps is acting as a rudimentary filter.

## 7.7 Environmental Noise

A principle design goal of Slocalization is accurate localization of a static tag in a static environment with static anchors. However, in many real-world scenarios, while the localization target may be stationary, the environment is not. Non-stationary environments will appear as noise in the CFR. As a baseline, in Figure 13a we capture the CFR noise for a static environment. We then consider the obvious environmental noise source for indoor spaces, namely people moving throughout the environment. In practice human beings do not move quickly in physical space, and Figure 13b shows that the simple addition of a 50 Hz high-pass filter is able to remove most of the CFR noise created by people moving about the space. The next source of noise Slocalization must deal with is that emitted by ambient devices in the space. In Figure 13c we find that the fluorescent lighting in our office building emits significant noise not successfully filtered by the 50 Hz filter added for removing human motion. Raising this filter to 150 Hz successfully removes the noise introduced by the lighting, facilitating Slocalization. It is in Slocalization's interest to keep this filter value as low as possible. The primary energy cost for the tag is throwing the antenna load switch, thus the lower the switching frequency, the lower the tag's active power draw. In practice we have not found other significant interference sources above 150 Hz testing in both a traditional office setting and a home environment. We set the tag oscillation frequency to 256 Hz to balance active power draw and detectability.

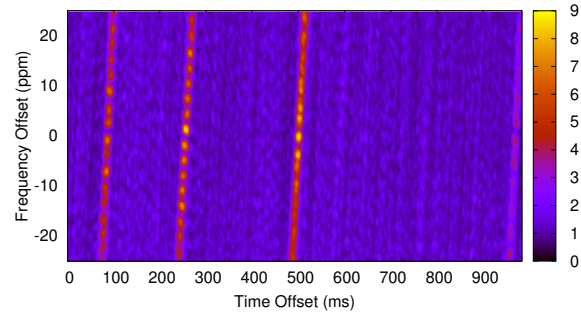

**Figure 14: Searching for Tags in Multi-User Settings.** To generate the backscatter tag CIR, the time offset and frequency offset of the backscatter modulation sequence must be determined. In the case of PN-coded backscatter transmissions, this search space can be quite large. This shows the resulting correlation search space for a PN code of length 63 transmitted with a period 983 ms. Three tags can be observed after an exhaustive search is performed. The peak values for each tag are used to accurately correlate and reconstruct their corresponding backscatter CIRs. A 63 bit PN allows concurrent localization of 63 uncoordinated tags.

## 7.8 Multiple Tags

The Slocalization design includes PN codes to allow the anchor infrastructure to distinguish multiple tags. Figure 14 places three concurrently transmitting Slocalization tags in the environment. The Slocalization system is able to cleanly distinguish each tag and localize it independently of the others.

## 7.9 Microbenchmarks

Our prototype tag—including the photovoltaic harvesting frontend—measures $5.5 \times 1.5\,\text{cm}$ and weighs just 3.5 g. The tag draws 406 μW while the microcontroller is active and 522 nW while it is in standby. Driving a worst-case constantly switching 0-1 signal through eight 74LVC595A [37] 8-bit shift registers at 512 Hz draws 277 nW, for a combined 800 nW during steady state operation.

## 8 DISCUSSION

With Slocalization, we have demonstrated the viability of UWB backscatter and shown the feasibility of localizing microwatt tags with decimeter-level accuracy. Before closing, we explore how much further Slocalization could go, and what could be done to make it faster (or equivalently cover larger areas)? Could Slocalization be used to localize something smaller than a grain of rice?

### 8.1 Speeding Up Slocalization

While Slocalization's performance is acceptable for a large array of devices and applications, there are numerous enhancements that could improve SNR, thus reducing required integration time, and accelerating localization. The RF frontends we employ exhibit an approximately 12 dB noise figure across the range of utilized frequencies. This offers the potential for improvement with the addition of a low-noise amplifier at each anchor receive antenna. Currently, Slocalization uses omnidirectional antennas to maximize anchor placement flexibility. WiTrack employs directional antennas following the argument that the most likely deployment scenario is "in the walls." The same is likely true for Slocalization in many cases. Replacing the current omnidirectional antennas [3] with directional UWB antennas [1] could realize at least 5 dB of gain. The instantaneous bandwidth measured at each step is smaller than that attainable with the radio hardware utilized, as the gigabit Ethernet communication used by the USRP N210 bottlenecks throughput. Larger instantaneous bandwidth could be attained by averaging on the FPGA fabric, lowering the necessary Ethernet bandwidth and therefore increasing the sweep rate and attainable update rate given the same specifications.

### 8.2 Scaling Up Slocalization

The frequency stability and precision requirements outlined in Section 4.2 for the normal operation of Slocalization are the same as the requirements needed to support frequency division. Coupling frequency division with the PN code division shown in this paper results in a multiplicative increase in the number of tags that can be simultaneously localized. This could be further enhanced by exploiting the stationary nature of tags. Over a long window of time (say, hourly) a tag could rotate through PN codes. The localization engine would collect the order of PN sequences over time at the same location to provide another dimension for distinguishing tags.

### 8.3 Shoring Up Slocalization

Prior localization schemes have consistently demonstrated that even just one or two range estimates beyond the minimum significantly improve localization performance, especially in the long tail [20, 23]. In a bistatic configuration, the number of channel soundings scales linearly with the number of anchors, as every other anchor can listen while one anchor is transmitting, enabling efficient capture of many range estimates in parallel.

Our prior work has also demonstrated that deploying multiple antennas at each anchor can help ameliorate orientation issues, cross-polarization, or nulls [21]. The current USRP N210 anchor cannot record the signal received at three antennas in parallel, however, thus exploiting antenna diversity with the current system would require further reduction in update rate.

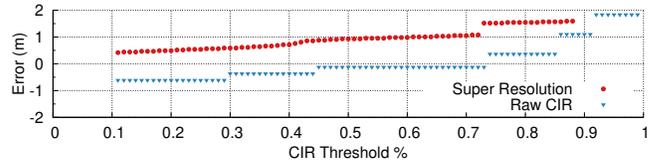

**Figure 15: ToA Estimation Error.** Using threshold-based estimation requires that the chosen threshold lie above the noise in the CIR, otherwise the arrival time estimation will strike noise far too early rather than the desired arrival peak (resulting in range estimation error much greater than 1 m for CIR thresholds below 10% in this case). Simplistic super resolution methods, such as the interpolation from Slocalization's zero-padding of the CFR, can provide greater fidelity, but have limited impact and may be skewed by outliers. For well-integrated samples, the ToA estimation mechanism is likely one of the largest causes of error in Slocalization measurements.

### 8.4 Cleaning Up Slocalization ToA

Fixed thresholding is one of the simplest techniques for estimating arrival time, and can contribute inaccuracies, especially when CIR noise is less predictable [16]. Ideally, tag arrival would be a vertical pulse in the CIR. One of the fundamental advantages of using UWB signals for localization is the narrower, tighter pulse shape in the time domain, which enables better estimation of actual signal arrival time. Still, UWB pulses have shape, and in a clean channel it is the leading edge of the pulse that captures the actual arrival time, not the peak. Figure 15 shows how increasing the CIR threshold affects the estimated distance as the arrival estimate moves up the peak.

The zero-padding of the CFR during Slocalization processing is a very basic form of super resolution, affording the finer-resolution steps in Figure 15. In RFind, Ma et al. observe that simply estimating ToA from the CIR discards valuable phase information [31]. Leveraging this, they develop a new super resolution technique that affords sub-centimeter accuracy. With the even greater bandwidth available to Slocalization, and provided that Slocalization as shown can achieve 0.07 m accuracy on its own for a given measurement, combining these techniques could theoretically realize sub-microwatt, sub-millimeter room localization.

### 8.5 Scaling Down Slocalization

Recently, there has been growing interest and initial demonstrations of viable millimeter-scale systems [24, 35, 42], so-called "smart dust." Whole room millimeter-accurate localization addresses a key deployment challenge for systems less than a millimeter in size.

Fundamentally, a Slocalization tag requires very little: a stable clock source, a shift register, and a variable impedance antenna element. Leveraging recent advances in near threshold circuit and oscillator designs, these components could be realized with a power budget on the order of nanowatts [7]. As nodes shrink, however, their physical antennas necessarily shrink as well, significantly reducing gain. Electrically small UWB antennas are still an active area of research, but the smallest antennas yielding high efficiency (near 0 dBi) are around 1 cm across [49]. A recent effort to optimize antennas for mm-scale nodes showed that narrowband mm-scale antennas realize gains of around -15 dBi within the Slocalization frequency range [6]. Assuming a similar correlation to achievable

UWB antenna gain along with the doubling in path loss due to the backscatter link, the Slocalization system would be required to realize another 30 dB of gain. This 30 dB of additional gain makes the integration times required for the current system intractable, but higher instantaneous bandwidth (up to 49× = 17 dB) and lower noise figure (12 dB) would almost completely make up the difference.

## 9   CONCLUSIONS

We show that by using ultra wideband backscatter, it is possible to realize both high accuracy localization and low energy operation, demonstrating long-range, decimeter-accurate positioning on a sub-microwatt power budget without requiring any tag or environmental motion. This is enabled by embracing the localization of stationary devices, facilitating the long-term integration of the channel to recover signals far below the noise floor. Slocalization lowers the burden of localization for the long tail of everyday objects, inviting a future where location information is ubiquitous.

## 10   ACKNOWLEDGMENTS

This work was supported in part by the CONIX Research Center, one of six centers in JUMP, a Semiconductor Research Corporation (SRC) program sponsored by DARPA, and in part by Terraswarm, an SRC program sponsored by MARCO and DARPA. We would like to thank our reviewers for their insightful comments and our shepherd, Andrew Markham, for support and guidance towards evaluations that greatly strengthened the final result. Finally, Slocalization could not have been a success without the support of Lab11, especially Joshua Adkins, Branden Ghena, Neal Jackson, and Noah Klugman.

## REFERENCES

[1] A. M. Abbosh and M. E. Bialkowski 2007. A UWB directional antenna for microwave imaging applications. In *2007 IEEE Antennas and Propagation Society International Symposium*.
[2] F. Adib, Z. Kabelac, D. Katabi, and R. C. Miller 2014. 3D Tracking via Body Radio Reflections *(NSDI'14)*.
[3] R. Azim, M. T. Islam, and N. Misran, Compact Tapered-Shape Slot Antenna for UWB Applications. *IEEE Antennas and Wireless Propagation Letters* 10 (2011).
[4] B. Campbell and P. Dutta 2014. An Energy-harvesting Sensor Architecture and Toolkit for Building Monitoring and Event Detection *(BuildSys'14)*.
[5] V. Chawla and D. S. Ha, An overview of passive RFID *(IEEE-COMM'07)*.
[6] Y. Chen, N. Chiotellis, L. X. Chuo, C. Pfeiffer, Y. Shi, R. G. Dreslinski, A. Grbic, T. Mudge, D. D. Wentzloff, D. Blaauw, and H. S. Kim, Energy-Autonomous Wireless Communication for Millimeter-Scale Internet-of-Things Sensor Nodes *(J-SAC'16)*.
[7] M. Choi, T. Jang, S. Bang, Y. Shi, D. Blaauw, and D. Sylvester, A 110 nW Resistive Frequency Locked On-Chip Oscillator with 34.3 ppm/°C Temperature Stability for System-on-Chip Designs. *JSSC* 51, 9 (2016).
[8] D. Dardari, R. D'Errico, C. Roblin, A. Sibille, and M. Z. Win, Ultrawide Bandwidth RFID: The Next Generation? *Proc. IEEE* 98, 9 (2010).
[9] S. DeBruin, B. Campbell, and P. Dutta 2013. Monjolo: An Energy-harvesting Energy Meter Architecture *(SenSys'13)*.
[10] R. D'Errico, M. Bottazzi, F. Natali, E. Savioli, S. Bartoletti, A. Conti, D. Dardari, N. Decarli, F. Guidi, F. Dehmas, and others 2012. An UWB-UHF semi-passive RFID system for localization and tracking applications *(RFID-TA'12)*.
[11] Energizer. CR2032 Datasheet. (2017).
[12] R. Exel 2012. Clock synchronization in IEEE 802.11 wireless LANs using physical layer timestamps *(ISPCS'12)*.
[13] R. Exel, T. Bigler, and T. Sauter, Asymmetry Mitigation in IEEE 802.3 Ethernet for High-Accuracy Clock Synchronization *(IEEE-TIM'14)*.
[14] Federal Communications Commission 2002. *First Report and Order 02-48*. Technical Report. Federal Communications Commission.
[15] S. Gezici, Z. Tian, G. B. Giannakis, H. Kobayashi, A. F. Molisch, H. V. Poor, and Z. Sahinoglu, Localization via Ultra-Wideband Radios: A Look at Positioning Aspects for Future Sensor Networks *(IEEE-SPM'05)*.
[16] I. Guvenc and Z. Sahinoglu 2005. Threshold-based TOA estimation for impulse radio UWB systems. In *2005 IEEE International Conference on Ultra-Wideband*.

[17] V. Heiries, K. Belmkaddem, F. Dehmas, B. Denis, L. Ouvry, and R. D'Errico 2011. UWB backscattering system for passive RFID tag ranging and tracking *(ICUWM'11)*.
[18] W. Hirt 2007. The European UWB Radio Regulatory and Standards Framework: Overview and Implications *(ICUWB'07)*.
[19] P. Hu, P. Zhang, and D. Ganesan 2015. Laissez-Faire: Fully Asymmetric Backscatter Communication *(SIGCOMM'15)*.
[20] B. Kempke, P. Pannuto, B. Campbell, and P. Dutta 2016. SurePoint: Exploiting Ultra Wideband Flooding and Diversity to Provide Robust, Scalable, High-Fidelity Indoor Localization *(SenSys'16)*.
[21] B. Kempke, P. Pannuto, and P. Dutta 2016. Harmonium: Asymmetric, Bandstitched UWB for Fast, Accurate, and Robust Indoor Localization *(IPSN'16)*.
[22] S. Kim, R. Vyas, J. Bito, K. Niotaki, A. Collado, A. Georgiadis, and M. M. Tentzeris, Ambient RF Energy-Harvesting Technologies for Self-Sustainable Standalone Wireless Sensor Platforms. *Proc. IEEE* 102, 11 (2014).
[23] Y.-S. Kuo, P. Pannuto, K.-J. Hsiao, and P. Dutta 2014. Luxapose: Indoor Positioning with Mobile Phones and Visible Light *(MobiCom'14)*.
[24] Y. Lee, S. Bang, I. Lee, Y. Kim, G. Kim, M. H. Ghaed, P. Pannuto, and others, A Modular 1 mm³ Die-Stacked Sensing Platform with Low Power I²C Inter-die Communication and Multi-Modal Energy Harvesting. *JSSC* 48, 1 (2013).
[25] V. Liu, A. Parks, V. Talla, S. Gollakota, D. Wetherall, and J. R. Smith 2013. Ambient Backscatter: Wireless Communication out of Thin Air *(SIGCOMM'13)*.
[26] P. Loschmidt, R. Exel, and G. Gaderer, Highly accurate timestamping for ethernet-based clock synchronization. *Journal of Comp. Networks and Comm.* (2012).
[27] D. Lymberopoulos, J. Liu, X. Yang, A. Naguib, A. Rowe, N. Trigoni, and N. Moayeri. Microsoft Localization Competition. (2015).
[28] D. Lymberopoulos, J. Liu, Y. Zhang, P. Dutta, Y. Xue, and A. Rowe. Microsoft Localization Competition. (2016).
[29] Y. Ma, X. Hui, and E. C. Kan 2016. 3D Real-time Indoor Localization via Broadband Nonlinear Backscatter in Passive Devices with cm Precision *(MobiCom'16)*.
[30] Y. Ma, N. Selby, and F. Adib 2017. Drone Relays for Battery-Free Networks *(SIGCOMM'17)*.
[31] Y. Ma, N. Selby, and F. Adib 2017. Minding the Billions: Ultra-wideband Localization for Deployed RFID Tags *(MobiCom'17)*.
[32] Maxim Integrated 2002. *Application Note 505 – Lithium Coin-Cell Batteries: Predicting an Application Lifetime*. Technical Report.
[33] C. McElroy, D. Neirynck, and M. McLaughlin 2014. Comparison of wireless clock synchronization algorithms for indoor location systems *(ICC'14)*.
[34] A. F. Molisch, D. Cassioli, C. C. Chong, S. Emami, A. Fort, B. Kannan, J. Karedal, J. Kunisch, H. G. Schantz, K. Siwiak, and M. Z. Win, A Comprehensive Standardized Model for Ultrawideband Propagation Channels *(IEEE-TAP'06)*.
[35] T. Nakagawa, M. Miyazaki, G. Ono, R. Fujiwara, T. Norimatsu, T. Terada, A. Maeki, Y. Ogata, S. Kobayashi, N. Koshizuka, and K. Sakamura 2008. 1-cc computer using UWB-IR for wireless sensor network *(ASPDAC'08)*.
[36] A. Nemmaluri, M. D. Corner, and P. Shenoy 2008. Sherlock: Automatically Locating Objects for Humans *(MobiSys'08)*.
[37] Nexperia. 74LVC595A Datasheet. (2014).
[38] A. N. Parks, A. P. Sample, Y. Zhao, and J. R. Smith 2013. A wireless sensing platform utilizing ambient RF energy *(BioWireleSS'13)*.
[39] RadiantRFID. Virtual Asset Tracker. (2017).
[40] R. M. Richardson. Remotely actuated radio frequency powered devices. (1963). US Patent 3,098,971.
[41] A. P. Sample, D. J. Yeager, P. S. Powledge, A. V. Mamishev, and J. R. Smith, Design of an RFID-Based Battery-Free Programmable Sensing Platform. *IEEE TIM* 57, 11 (2008).
[42] Y. Shapira. Tel Aviv U Review. english.tau.ac.il/sites/default/files/media_server/TAU%20Review%202008-09.pdf. (2008). Smart Dew.
[43] C. Swedberg. Children's of Alabama Expands RTLS Deployment for Asset Tracking. (2013). rfidjournal.com/articles/view?10479/2.
[44] V. Talla and J. R. Smith 2013. Hybrid analog-digital backscatter: A new approach for battery-free sensing *(RFID'13)*.
[45] C. Walton. Portable radio frequency emitting identifier. (1983). US Patent 4,384,288.
[46] J. Wang, D. Vasisht, and D. Katabi 2014. RF-IDraw: Virtual Touch Screen in the Air Using RF Signals *(SIGCOMM'14)*.
[47] WiseTrack. (2017). wisetrack.com/rfid-asset-tracking-equipment.
[48] L. Yang, Y. Chen, X.-Y. Li, C. Xiao, M. Li, and Y. Liu 2014. Tagoram: Real-time Tracking of Mobile RFID Tags to High Precision Using COTS Devices *(MobiCom'14)*.
[49] T. Yang, W. A. Davis, and W. L. Stutzman 2008. The design of ultra-wideband antennas with performance close to the fundamental limit *(URSI'08)*.
[50] H. Zhang, J. Gummeson, B. Ransford, and K. Fu, Moo: A batteryless computational RFID and sensing platform. *University of Massachusetts Amherst, Tech. Rep* (2011).
[51] P. Zhang, J. Gummeson, and D. Ganesan 2012. BLINK: A High Throughput Link Layer for Backscatter Communication *(MobiSys'12)*.
[52] L. Zhou, G. Li, Z. Zheng, and X. Yang 2014. TOA Estimation with Cross Correlation-Based Music Algorithm for Wireless Location *(CSNT'14)*.